%% file: aerospace2022-rce-for-mbse.tex
\documentclass{llncs} 
\usepackage{fullpage}

\usepackage{cite}
\usepackage{amsmath,amssymb,amsfonts}
\usepackage{algorithmic}
\usepackage{graphicx}
\usepackage{textcomp}
\usepackage{xcolor}
\usepackage{xspace}
\usepackage{caption}
\usepackage{subcaption}
\usepackage[inline]{enumitem}
\usepackage{hyperref}
\usepackage{cleveref}
\usepackage{nameref}

\usepackage{tikz}
\usetikzlibrary{calc}

\newcommand{\Cologne}{Cologne}
\newcommand{\SysML}{SysML\xspace}
\newcommand{\MBSE}{MBSE\xspace}
\newcommand{\DLR}{\text{DLR}\xspace}
\newcommand{\RCE}{\textsc{RCE}\xspace}
\newcommand{\CPACS}{\textsc{CPACS}\xspace}
\newcommand{\mydef}[1]{\emph{#1}}

\begin{document}

\title{Orchestrating Tool Chains for Model-based Systems Engineering with RCE}

\author{
  Jan Flink \and
  Robert Mischke \and
  Kathrin Schaffert \and
  Dominik Schneider \and
  Alexander Weinert%
  \thanks{Authors are listed in alphabetical order}
}
\institute{
  Institute for Software Technology\\
  German Aerospace Center (DLR) \\
  51147 \Cologne, Germany\\
  \email{firstname.lastname@dlr.de}
}

\maketitle

\begin{abstract}
  When using multiple software tools to analyze, visualize, or optimize models in \MBSE, it is often tedious and error-prone to manually coordinate the execution of these tools and to retain their respective input and output data for later analysis.
  Since such tools often require expertise in their usage as well as diverse run-time environments, it is not straightforward to orchestrate their execution via off-the-shelf software tools.
  We present \RCE, an application developed at the German Aerospace Center that supports engineers in developing and orchestrating the execution of complex tool chains.
  This application is used in numerous research and development projects in diverse domains and enables and simplifies the creation, analysis, and optimization of models.
\end{abstract}

\section{Introduction}
\label{sec:introduction}

In classical engineering, work processes are organized around textual documents.
Such documents contain, e.g., the specification of requirements, preliminary designs of the system under construction, or test reports of prototypes.
These documents are written in natural language and are prone to accidental incompleteness, ambiguity, or conflicting definitions.

To alleviate these issues, \MBSE~(\mydef{Model-based Systems Engineering}) is applied in more and more engineering projects.
In contrast to classical engineering principles, the processes of \MBSE focus on the production and investigation of formal models instead of documents~\cite{INCOSE2007}.

There exist numerous processes, techniques~\cite{Estefan2008,MadniSievers2018}, and tools~\cite{RashidAnwarKhan2015} that engineers can use to construct such a model in a reliable, timely, and precise way.
Once an initial version of the model is produced, engineers may use the model to evaluate the behavior of the system in specific scenarios.
During the development process, the model will go through multiple iterations as the understanding of the system improves and the fidelity of the modeling increases.

The specification language used for describing the model depends strongly on the domain of engineering.
One example of such a language is \SysML~(\mydef{Systems Modeling Language})~\cite{OMG2019}, which describes a system via a set of diagrams.
Each diagram provides a different view on the system, e.g., on the structure of its components or on the communication between them in a given usage scenario.
There also exist more application-specific modeling languages that allow for greater ease of modeling.
One example in the domain of aircraft design is \CPACS~(\mydef{Common Parametric Aircraft Configuration Schema})~\cite{AlderMoerlandJepsenEtAl2020}.

Independently of the modeling language used, most activities of \MBSE involve software tools that ``understand'' the constructed model and perform some operation with it.
These operations may involve, e.g., the extension of the model with some characteristics, the evaluation of the model's performance in some scenario, or its optimization with respect to some property.

We present \RCE (\url{https://rcenvironment.de/}), a freely available open-source application for designing, executing, and orchestrating automated distributed tool chains.
\RCE is being developed at \DLR (German Aerospace Center) at the Institute for Software Technology.
It satisfies multiple requirements for executing fully automated tool chains.
Among others, \RCE allows users to integrate almost arbitrary tools for simulation, analysis, and optimization into tool chains, to publish these integrated tools in a network, and to quickly adapt the design of tool chains to new requirements via a graphical interface.

The remainder of this paper is structured as follows:
After defining nomenclature in Section~\ref{sec:preliminaries}, we give an overview over the main features of \RCE in Section~\ref{sec:rce} and describe the design decisions and trade-offs made in the practical implementations of these features in Section~\ref{sec:practice}.
We subsequently highlight research and development projects in which \RCE has been successfully applied in Section~\ref{sec:use-cases} before discussing related work and future directions for development in Section~\ref{sec:related-work} and Section~\ref{sec:future-work}, respectively.

\section{Nomenclature and Preliminaries}
\label{sec:preliminaries}

In this section we introduce concepts underlying our work as well as the nomenclature that we use in the remainder of this work.

The major novelty of \MBSE in contrast to previous engineering processes is its focus on a common model of the system as the main artifact around which engineering processes are centered.
This model is defined in a formal language that is used by all process participants.
The availability of such a formal model allows for a greater support by \mydef{(software) tools} than for non-model-based engineering processes.
These tools support engineers in, e.g., analyzing, optimizing, or visualizing the model.~\cite{AdamsBohnhoffDalbeyEtAl2021,SiggelKleinertStollenwerkEtAl2019}

Parts of each of these activities may be implemented comprehensively in a single tool, such as the visualization of the complete model under construction.
In most cases, however, several tools are required to perform an activity, using the output of one tool as the input for one or more succeeding ones.
We say that the individual software tools form a \mydef{tool chain} in this case.

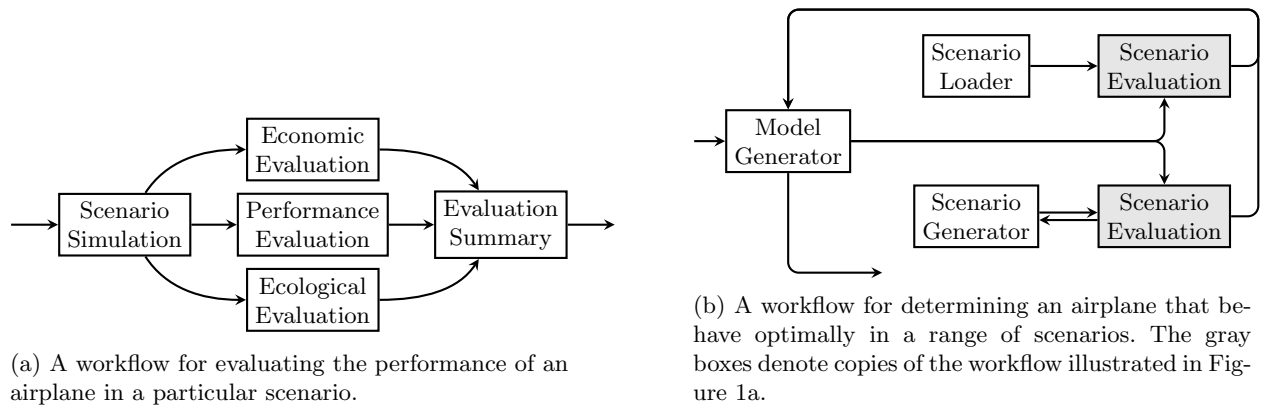
\begin{figure*}
  \centering
  \begin{subfigure}[b]{.45\textwidth}
    \input{images/workflow-concept-evaluation}
    \caption{A workflow for evaluating the performance of an airplane in a particular scenario.}
    \label{fig:preliminaries:workflow:evaluation}
  \end{subfigure}
  \hfill
  \begin{subfigure}[b]{.45\textwidth}
    \input{images/workflow-concept-main}
    \caption{A workflow for determining an airplane that behave optimally in a range of scenarios. The gray boxes denote copies of the workflow illustrated in Figure~\ref{fig:preliminaries:workflow:evaluation}.}
    \label{fig:preliminaries:workflow:main}
  \end{subfigure}
  \caption{A workflow for determining an optimal airplane.}
  \label{fig:preliminaries:workflow}
\end{figure*}

Consider, e.g., the evaluation of an airplane model in a given scenario.
First, the behavior of the model in the scenario needs to be simulated.
Afterwards, the collected data needs to be evaluated with respect to at least the performance of the airplane as well as its economic and ecological characteristics.
To simplify the comparison of different scenarios, these three evaluations must then be consolidated into a single evaluation.
Each step of this process may in practice be implemented in a distinct software tool.
We illustrate the individual steps and the relation between them in Figure~\ref{fig:preliminaries:workflow:evaluation}.

While simple tool chains may suffice for many engineering tasks, they may become arbitrarily complex and require iterated or nested execution.
Consider, e.g., the task of evaluating the behavior of the same airplane shape with different engines in diverse scenarios.
These scenarios may either be based on historical data, or they may be generated according to anticipated operational scenarios.
Given an initial airplane shape, for each candidate engine first a model will be constructed that combines the model of the airplane shape with a model of the engine.
The behavior of this combined model can then be simulated in multiple scenarios, where each simulation results in a comprehensive evaluation of the model's performance.
By iterating over each candidate engine, the performance of all engines can subsequently be compared.
We illustrate the relation between the individual steps in such an evaluation in Figure~\ref{fig:preliminaries:workflow:main}.

Tools in a workflow may be distributed among individual computers on a network.
In a typical engineering process, each tool in a workflow has an operator who is familiar with its operation and responsible for providing the service.
To execute a workflow, the operators of the initial tool(s) first execute these tool(s) before passing the results of this execution on to the operators of subsequent tools.
Passing of information among operators often happens via, e.g., shared network drives or via email.
The subsequent operators then execute their respective tools and again pass their outputs to subsequent operators.
The operators may also pre- or post-process the data they receive and produce using additional ``helper'' tools.

This non-standardized, manual process is time-consuming, tedious, and prone to errors.
It also puts the burden of manual bookkeeping on operators, without an effective way to assure accuracy or completeness.
For example, accidentally reusing an intermediate file from a previous, structurally similar run can lead to incorrect results while being very hard to detect.

Due to this, it is often cumbersome or even impossible to track down the source of any deviations or inconsistencies seen in the final output of a workflow.
Even more concerning is the potential for hidden, non-obvious errors, which can corrupt research results, and may go undetected indefinitely.

In the following sections, we present our software \RCE, which allows engineers to construct and automatically execute \mydef{workflows} comprising tools that are available via network connections.
These workflows comprise the major tools present in the workflow as well as additional ``helper'' tools that, e.g., control the flow of information or convert between data formats.
\RCE imposes few requirements on the software tools contained in the workflow and none on their packaging, i.e., it does not require software to be available as a single file or archive.
Thus, it greatly simplifies the construction and execution of workflows comprising a wide range of software, both off-the-shelf and custom-made.

During execution of a workflow, the tools remain on their respective machines, only the input and output data is moved around the network.
This allows tool operators and tool developers to retain full control over the version of the tool that they provide to other engineers.
Moreover, during execution of a workflow, \RCE retains both the inputs and outputs of each individual execution of a tool in a central database.
This enables engineers to easily trace unexpected result data to its original occurrence and to retain detailed information about the origin of simulation data.

\section{\RCE}
\label{sec:rce}

Targeting engineers' needs for integrating \MBSE methods into day-to-day-work, \RCE provides a framework for
\begin{enumerate*}[label=\alph*)]
	\item controlling the data flow of engineer tools' input and output,
	\item sharing tools across organizational borders in a secure manner, and
	\item automatically executing distributed workflows.
\end{enumerate*}
As of January 2022, \RCE supports multiple operating systems such as Microsoft Windows and Windows Server as well as various Linux distributions.
It is possible to run \RCE with a GUI (Graphical User Interface) or as terminal-only application which makes it suitable for use on both workstations and servers.
While GUI instances are mainly used to design and execute workflows and to subsequently access the result data, terminal instances are usually used as a tool server or network connection point.

In the remainder of this section we present a step-by-step introduction to \RCE from a user's perspective.
We start with the integration of a software tool into \RCE and conclude with the execution of a distributed workflow.
For a more technical description of \RCE's features and capabilities, we refer to work by Boden, Flink, F{\"o}rst, et al.~\cite{BodenFlinkFoerstEtAl2021}

Before executing a tool via \RCE, it first has to be \mydef{integrated} into \RCE, thus turning it into a \mydef{workflow component}.
To integrate a tool, \RCE requires it to be executable as a terminal application without further user interaction during execution. This means that tools can launch a graphical user interface as part of their operation when needed, but must not require any user interaction to complete their execution.
A tool can be integrated either manually via configuration files or with the ``Tool Integration Wizard'', which automatically creates these configuration files.
This allows users to simply integrate a tool with the aid of the wizard and to export the configuration files to a server installation afterwards.
As tools may be executed in heterogeneous environments, users can define commands for the tool's execution under Windows and Linux, making the configuration of an integrated tool independent of the operating system.
Part of the integration process is defining a set of \mydef{inputs} and \mydef{outputs} for the respective workflow component.
These inputs and outputs then can be accessed by the user-defined \mydef{pre-processing script}, \mydef{execution script}, and \mydef{post-processing scripts}.
These scripts adapt incoming and outgoing data for the tool itself and for the following components within the workflow.

\RCE provides multiple options for sharing integrated tools with other \RCE instances. 
Users can create networks of multiple \RCE instances for sharing access to components and distributed workflow execution. 
A connection between two \RCE instances can be either a normal connection in the local network, or it can be an encrypted \mydef{Uplink} connection.
At least one \RCE instance has to be configured as server instance for either incoming local network or Uplink connections.
To prevent unauthorized access, workflow components have to be shared explicitly for each component via the ``Component Publishing'' view.
Users can furthermore create arbitrary groups for sharing components.
Shared components are then visible for each connected \RCE instance only if the respective instances are part of the same authorization group.
We describe the authorization and security concept in detail in Section~\ref{sec:practice:authorization-tool-access}.

After integrating a tool and thus turning it into a workflow component, users can use shared and local components to construct a workflow via the graphical ``Workflow Editor''.
This editor is part of the GUI and the main view for editing workflows.
In addition to integrated components, \RCE includes a set of standard components for often-needed basic functionalities.
These components easily allow for, e.g., reading and writing files or controlling the data flow within the workflow.
Additionally, the users can execute their own Python scripts directly as part of the workflow, handle XML data, or create loops for evaluation or optimization purposes.
For the creation of optimization loops, \RCE includes the Dakota framework~\cite{AdamsBohnhoffDalbeyEtAl2021}.
Users may, however, opt to use other optimization frameworks.
It is possible to have multiple instances of the same component within a workflow.

\begin{figure*}
	\centering
	\includegraphics[width=15cm]{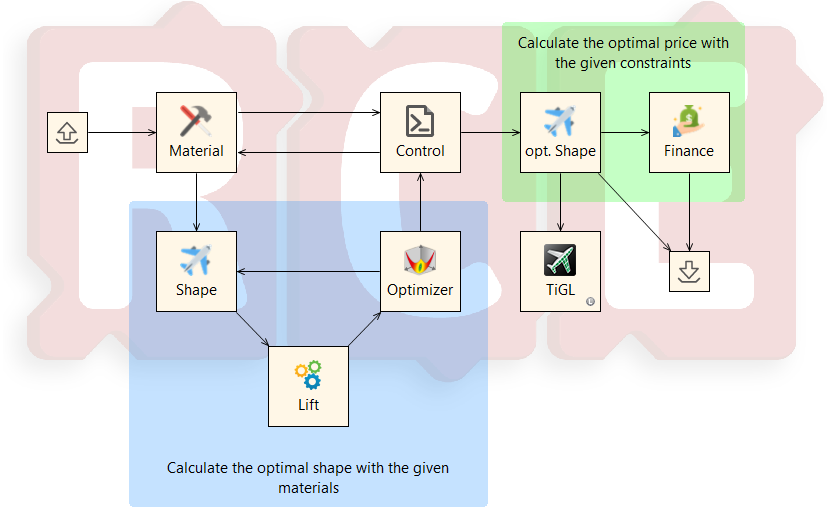}
  \caption{%
    A workflow in \RCE. %
    Originally produced by Boden, Flink, F{\"o}rst, et al.~\cite{BodenFlinkFoerstEtAl2021}, used with permission.%
  }
	\label{fig:rce-workflow}
\end{figure*}

The ``Workflow Editor'' provides utilities for workflow design, such as connecting the outputs of one component to the inputs of another component or creating labels for marking groups or important calculation steps. 
\Cref{fig:rce-workflow} shows a workflow within the ``Workflow Editor.''
The yellow boxes represent either an integrated or a standard component whereas the black arrows represent connections from outputs to inputs.
The blue and the green rectangle are the labels marking groups of closely related component instances.
It is possible to have arbitrarily many connections between a pair of tools.
For the sake of readability, connections with the same direction are displayed as a single arrow.

Besides the configuration of component publishing and network connections within their respective views, \RCE abstracts the complexity of shared components within the ``Workflow Editor'':
Local and shared integrated components are indistinguishable from one another during the workflow editing process.

After having constructed a workflow using both local and shared workflow components, the user can execute it.
When executing a workflow, the user has to determine an \RCE instance within the network which serves as \mydef{workflow controller}.
This workflow controller stores all intermediate execution data and orchestrates each component execution within the workflow.
This centralized design allows to disconnect \RCE instances during the execution of a workflow if they do not publish components included in the workflow.
Consider, e.g., a user who designs a workflow on their mobile computer.
They can designate a long-running server as the workflow controller for the execution of the workflow.
After starting the execution, the user can disconnect and connect their mobile computer at will without interrupting the workflow's execution. 

\RCE allows the users to integrate the same tool at multiple locations in the same network.
In that case, the users have to choose one of the multiple available integrations for its execution before starting the workflow.
After starting a workflow, \RCE executes all components which do not need input data and forwards their output data to the subsequent component in the workflow.
The workflow controller automatically receives and distributes the data to the respective host instances of each component via the network setup.
During the execution of a component representing an integrated tool, all of it's scripts, i.e, the pre-processing, execution, and post-processing script, are executed in order.
The workflow finishes when no component produces any further output.

All input and output data of all component executions of the workflow run are stored in the ``Data Management''.
Already during the execution of a workflow and after its completion, these data can be accessed by all \RCE instances connected to the workflow controller via the \RCE ``Data Management Browser'', which is included in the GUI of \RCE.
Thus, users can benefit from collaborative access to the result data.
Moreover, additional data of the workflow and component executions are stored in the data management and enable the workflow run to be monitored and evaluated.
This includes start and end times of executions and log outputs of user integrated tools.

\subsection{User Roles Interacting with \RCE}
\label{sec:user-roles}

As previously described, \RCE allows to integrate tools and share the respective components to design workflows to develop and analyze complex systems.
Due to the many different tasks \RCE covers, users interact with \RCE in a number of different roles.
Most commonly, we identify the roles of tool integrator, workflow designer, and network administrator in real-life projects.~\cite{BodenMischkeWeinertEtAl2020}
A tool integrator focuses on the integration process for the used tools, e.g., setting up the different execution scripts.
A workflow designer, in contrast, can ignore the technical details of the tools and concentrates their work on the semantic correctness of the overall design.
Finally, a network administrator does not need to take the workflow design nor the tool integration into account, as their priority is to set-up a secure and preferably simple \RCE network structure.
In addition to these common roles, users reported the additional roles of workflow executor and data analyst in their respective use-cases.

To support these different modes of interaction, \RCE offers specific sets of instruments and the possibility to adapt the GUI to each role's respective requirements.
The advantage of this approach is that each role can focus on their task improving the efficiency of the whole workflow development process.
Therefore, \RCE's GUI consists of multiple views where each view holds an instrument.
A set of views is called a \mydef{perspective}.
\RCE allows to work with multiple perspectives, enabling seamless switching between tasks and their respective instruments.

\subsection{Development Process of \RCE}

\RCE has emerged from earlier, more specialized research projects, and has been under development since 2006.
Today, it is an open-source project publicly available for download from \url{https://rcenvironment.de/} with its source code available from \url{https://github.com/rcenvironment} on GitHub.

All changes to the code base are tracked with a version control software, allowing to preserve the whole history of the project and to access and rebuild older releases.
The development cycle is oriented on semantic versioning~\cite{PrestonWerner2013}, meaning there are major, minor, and patch releases on a regular basis.
Each release undergoes a quality management procedure containing, among other things, continuous integration, automated tests, and manual testing~\cite{MischkeSchaffertSchneiderEtAl2022}.

Following semantic versioning, all releases within a major cycle are compatible with each other, which is especially important for interoperation within \RCE networks.
Concerning the loading of existing workflows, we aim to provide backwards compatibility, even for older major release cycles, for as long as possible.
Whenever the structure of workflow files needed to change in the past, we have continuously provided automated upgrade capabilities.

\section{Practical Considerations}
\label{sec:practice}

In this section we address practical aspects of constructing, operating, and using distributed tool-based workflows.
We connect these aspects to various design decisions of \RCE, explain the trade-offs resulting from these decisions, and explain how the operation of \RCE networks fits into the technical and administrative environment of real-world organizations.

\subsection{Providing Tools as Computational Services}
\label{sec:practice:tool-services}

From a purely model-based perspective, the tools within a tool chain can be viewed as black boxes which are executed in an abstract computing environment.
In practice, however, the location and environment where tools are installed and executed is highly relevant.

For instance, tools may require specific \emph{operating system and software environments}, e.g., certain operating systems or the presence of specific pre-installed versions of libraries.
Further, different locations allow for different degrees of \emph{technical monitoring, access control and IT management}, only some of which may be sufficient to meet organizational policies.
Moreover, especially for high-performance computing tools, knowing and selecting the exact \emph{properties of the underlying hardware} is important for optimal performance.
Additionally, the \emph{network performance} of an installation location is also relevant for large or frequent data transfers.
Also, the \emph{ease of deployment} of tool updates may also vary greatly between locations.
Finally, \emph{billing for the usage of underlying resources} (e.g., computing time or licenses) may also be required in some environments.
Conversely, the usage of the tool itself may also be subject to \emph{auditing and accounting} needs, especially when used by external partners.

For any combination of these reasons, both developers and administrators typically want to choose the proper locations for the execution of their tools on a case-by-case basis.
In contrast, distributing actual tool installations (i.e., binary files or scripts) over the network is generally not desirable, and often not permitted for security reasons.
Instead, it is much more useful to publish and consume tools as abstract computational services, while executing them in these explicitly chosen locations.
Consequently, \RCE's fundamental approach to the distributed execution of multi-tool workflows is to provide the execution of pre-installed tools as remote-callable services.

While this service-based approach offers many advantages, it is not completely without downsides.
The most important one is that it is not possible to transparently move the execution of a tool to where the data to be processed resides, or where the generated output data is needed.
While mostly irrelevant for small volumes of data, this can become a performance issue when large amounts of data must be processed or generated.

While \RCE cannot, and is not meant to, solve this installation problem automatically, its flexible network structure allows users to develop solutions for these scenarios.
As the same tool can be provided from any number of locations, any tool that regularly processes large amounts of data can be installed close to the source or the receiver of that data.
By also publishing this additional installation within the \RCE network, this location becomes available as a selectable option when starting a workflow involving that tool.
By then choosing the appropriate tool locations at the start of the workflow, users can reduce the performance impact of large data transfers.

\subsection{Authorization of Tool Access}
\label{sec:practice:authorization-tool-access}

At the time of its inception, \RCE was built for closely-tied working groups with no need for access control on tools or data.
Nowadays, however, it is being deployed in multidisciplinary environments involving different projects, organizational units, and increasingly, even multiple organizations.
While these setups are protected from outside access using standard IT security measures, there are still valid reasons for further controlling the access to published tools.

For instance, tools frequently represent novel scientific or engineering approaches, which may not have been published yet.
While its author(s) may trust selected colleagues with access to it (e.g., for testing and evaluation), they may not want to provide it to a wider audience yet.

Tools may also be installed in a location where they can use confidential code or data for their computation.
For example, an industry partner may allow academic partners to simulate certain scenarios using proprietary technical datasets, without sharing that data directly.
Given organization-wide and unlimited access, however, a malicious actor could potentially craft specialized workflows to reverse engineer parts of these datasets.
To prevent this, such partners need an option to restrict these simulations to a minimal, well-known, and trusted set of users.

As an additional example, tools may also use limited or costly computational resources.
In that case, it is desirable to restrict the tool's usage to authorized members for proper billing, or to avoid unexpected cost overruns.

Starting with \RCE 9, \RCE offers a limited security mechanism to address these needs.
This mechanism consists of a group-based concept, with each group being defined by ownership of a shared secret cryptographical key.
From a user's perspective, joining a group is similar to accessing a password-protected area.
Every user can publish tools for any number of groups they are a member of, and consume all tools that are published for at least one group they are a member of.
If a tool is considered generally useful, and free from the restrictions mentioned above, \RCE still offers the option of publishing it to anybody within the same \RCE network.

Unlike a standard login-based approach, \RCE's access control is decentralized, i.e., there is no central server for checking a user's authentication.
Instead, any user can create a new group at any time, and invite other users to join it.
These groups are implicitly disposed when they are not used for sharing tools anymore.
This approach was chosen to match \RCE's focus on flexibility and bottom-up cooperation, and to avoid the overhead of a static authentication infrastructure.
We discuss a future evolution of this system in Section~\ref{sec:future-work:user-management}.

\subsection{Iterative Development and Updating of Tools}

Especially in research, frequent development, iteration, and experimentation regarding tools within a workflow is common.
These tool updates may involve new algorithmic approaches being tested, new features added, or issues processing particular input data being fixed.

\RCE is well-suited to this iterative work due to several properties.
Firstly, any tool that is updated in its execution location is immediately available in the new version to all users within the tool network.
Secondly, especially in early development, it is possible for a tool's author to publish a current version directly from their development machine.
While execution is usually not as performant as running it on a server or cluster, the rapid iteration and debugging possibilities this allows are often a great advantage.
Lastly, the deterministic and structured approach of \RCE's workflow execution makes it easy to re-run existing workflows with new versions of a tool.
This can be both used to validate pre-existing behavior when no changes are expected, and also to immediately verify the effect of improvements within the workflow's context.

\subsection{Cross-Organizational Tool and Workflow Publication}
\label{sec:practice:cross-organizational}

\RCE is widely used for collaboration between organizational units within larger organizations, e.g., multiple institutes of a single research center, each of which is specialized on different parts of the domain space.
A typical setup for this is a multidisciplinary workflow comprising tools published by these organizational units, thereby contributing to the overall workflow.

A natural expansion of this scenario is the integration of tools provided by domain experts located in disjoint organizations.
For example, industry partners with access to physical test sites or prototype designs and academic partners exploring new algorithms and methods may mutually benefit from combining their tools.

While this change regarding tool locations is virtually irrelevant from the MBSE domain perspective, it has profound technical, administrative, legal, and security implications.
From a technical standpoint, one major challenge is having all involved parties agree on a common protocol for exchanging access to the relevant tools.
Administratively, one important aspect is explicit control of which tool services should be made available to which partners, and potentially reviewing their usage.
Legally, both research and industry organizations are acutely aware of the need to protect intellectual property, especially in the form of tool code or data.
Lastly, IT security is of paramount importance, both as a requirement in and by itself, as well as providing the foundation for these legal and administrative needs.

As \RCE was not originally designed with cross-organization workflows in mind, adopting its standard network features to this would be difficult.
Instead, we opted to design and implement the entirely new, security-first network protocol named \mydef{Uplink}.
It is publicly available in \RCE 10 as an experimental feature and is currently being tested by several project groups.

Instead of attempting to retroactively restrict operations within the standard \RCE network, its design philosophy is the opposite.
Starting from scratch, only a restricted and safe set of operations is explicitly provided as needed.
In its current form, it supports announcing tool services and consuming them, with the auxiliary feature of transferring tool documentation to users.
In contrast, accessing the data management of \RCE instances in other organizations or starting remote workflows is explicitly forbidden.
While this is -- by design -- more restrictive than a local network, it allows fully automated integration of cross-organization tools into local \MBSE workflows.

\section{Concrete Use Cases}
\label{sec:use-cases}

In the previous section we have given an overview over the design decisions taken when developing \RCE and the trade-offs involved in this.
In this section we discuss real-life research projects using \RCE.

Many users already rely on \RCE in their daily work and appreciate its capabilities.~\cite{NoedingBertsch2021,PapanikolaouHarriesHooijmansEtAl2020,NiklassDzikusSwaidEtAl2020,CruzFregnaniMattosEtAl2021}
In addition to \DLR-external projects, there are many research projects at \DLR itself in which \RCE is used that contain concrete use cases.
These use cases are in application domains covering both the domain of aerospace as well as in the domains of traffic and energy.

We now highlight two projects actively using \RCE that implement \MBSE, namely JETSCREEN and EXACT.
Due to its nature as freely distributed open-source software, we omit a complete overview of all projects in which \RCE is involved.
Beyond that, Boden, Flink, F{\"o}rst, et al.~\cite{BodenFlinkFoerstEtAl2021} present further projects that use RCE.

\subsection{JETSCREEN}
\label{sec:use-cases:jetscreen}

The project JETSCREEN (JET Fuel SCREENing and Optimization) aims to support the development of sustainable aviation fuels and reduce climate impact of aviation.
To do so, it implements a model and a simulation-based process to generate key indicators out of a detailed composition of a candidate fuel.
These key indicators then support the fuel approval process.
The stated objectives of this project are ``to develop a screening and optimization platform, which integrates distributed design tools and generic experiments to assess the risks and benefits of alternative fuels, and to optimize alternative fuels for a maximum energy per kilogram of fuel and a reduction of pollutants emissions.''~\cite{Rauch2020}

To achieve this goal, project members implemented a cross-organizational \RCE network.
In this network, the individual disciplinary predictive and simulation software tools of the project partners were integrated into \RCE and made accessible to other participants.
Based on this infrastructure, \RCE workflows were developed which enable the automatic, distributed execution of the entire simulation processes.
These workflows include all the competencies of the project partners that are necessary for the evaluation of fuels.
To support the possibility of cross-company networking, the Uplink capabilities of \RCE were further developed as part of this project.

\subsection{EXACT}
\label{sec:use-cases:exact}

The currently ongoing \DLR-internal project EXACT (Exploration of Electric Aircraft Concepts and Technologies) has the overall aim of evaluating the potential of climate-neutral aircraft on a system level approach.
This evaluation includes both operational as well as social aspects. 
The \DLR research divisions of aviation, energy, space and transport are participating in this project.
In doing so, they, among other things, provide and extend their domain-specific simulation software tools.
As part of the project, various aircraft concepts are designed taking into account various factors such as energy carrier or seat-class.
In this use case, \RCE is applied to enable the fast and flexible integration of new technologies and disciplines into multidisciplinary dimensioning processes.
These iterative processes include several disciplinary, semi-empirical and physics-based models with low to medium fidelity.~\cite{SilberhornHartmannEtAl2020}

For this purpose, a \DLR-internal \RCE network was set up and the simulation tools were integrated into \RCE.
This means that all tools and models from the project partners are accessible within the integration framework for setting up the design tool chain.
In the context of this project, the ability of \RCE to simply integrate an \RCE workflow itself as a workflow component including the possibility to make it available in the network for reuse is being extended.
In particular, a GUI for this purpose is being developed.

\section{Related Work}
\label{sec:related-work}

The application that is closest to \RCE in terms of features it offers is ModelCenter, developed by Phoenix Integration~\cite{ModelCenter}.
Similarly to \RCE, this application allows users to combine their purpose-built software tools into a workflow and to execute that workflow automatically.
Model Center is proprietary software and not freely available.
\RCE, in contrast, is developed as open-source software and freely available under a permissive license.

Moreover, Apache Nifi~\cite{Nifi} allows its users to construct data pipelines.
Such a pipeline consists of individual processors that process incoming data and pass their output to one or more succeeding processors.
These pipelines are conceptually similar to the workflows that can be implemented in \RCE.
Nifi, however, focuses on indefinitely-running pipelines that continually process incoming data from a variety of sources.
The focus of \RCE, in contrast, lies on the implementation of long-running workflows that process a batch of data in a single execution.

Another project similar to \RCE is the Air Traffic Management (ATM) Test Bed by NASA~\cite{ChanBarmoreKiblerEtAl2018}.
This application allows users to simulate and experiment with concepts for the management of air traffic.
Similarly to \RCE, it features a graphical canvas on which users can arrange and connect individual scenario generation and simulation tools.
In contrast to \RCE, however, the ATM Test Bed specifically targets the domain of air traffic management.
It is, to the best of our knowledge, not well-suited to, e.g., the simulation of individual planes or the application in other domains such as the simulation of spacecraft or energy systems.

\section{Future Work}
\label{sec:future-work}

\RCE is under active development as part of multiple research and engineering projects.
A common theme of most ongoing work is maintaining the stability and further improving the efficiency and user-friendliness of \RCE.
Additionally, we are continuously monitoring technical and funcional needs, and support users in deploying \RCE for new use cases.
In this section, we highlight some of the main topics for the future development of \RCE.

\subsection{Decentralized Data Management}
\label{sec:future-work:data-management}

One major strength of \RCE is its support for automatically storing detailed input, intermediate, and output data for each executed workflow.
Each data transfer between integrated tools is recorded for later inspection.

While the execution of the individual tools is decentralized, this data archiving is centralized on the \RCE instance controlling the workflow execution.
This setup guarantees that all archived data is consistent and cannot be partitioned due to later unavailability of other instances.
The drawback of this approach, however, is the performance overhead of transferring all intermediate artifacts to a central server, which also limits its scalability.
Additionally, this centralized data retention can be problematic for data which is affected by special intellectual property regulations.

For future work, we are investigating the usefulness of decentralized data storage mechanisms for improved execution performance and scalability.
Depending on project needs, special provisions for more fine-granular access control to proprietary data may also be added.

\subsection{Provenance Recording}
\label{sec:future-work:provenance}

Furthermore, while the data management of \RCE stores all input and output data for each execution of the components of a workflow, it only retains limited information about the environment in which each component was executed.
The data management also does not currently store the explicit relation between the output of one component and the input of the subsequent component.
Instead, it only retains a copy of the executed workflow from which the user can reconstruct the input-output relations between component executions.

Explicitly storing such implicit information about the provenance of data would greatly assist users in developing and debugging workflows.
It would additionally allow for easier auditability and replicability of workflow execution.
We are currently working on a prototypical implementation of provenance recording in \RCE using the W3C PROV standard and provenance templates~\cite{MoreauMissierEtAl2013,MoreauBatlajeryHuynhEtAl2018}.

\subsection{Footprint Reduction, APIs, and Ecosystem Integration}
\label{sec:future-work:ecosystem}

Currently, \RCE is packaged as a single software application that can be used in all contexts discussed in this work.
This includes interactive work via its GUI, deployment as a tool publishing and workflow coordination platform, as an intermediate communication relay, or for batch execution by external scripts.
While this single release package is flexible enough to be used in all of these roles, it could be optimized by providing more specialized packages for each.
Especially for usage in containerized or cloud environments, reducing its deployment footprint by omitting client-side features would be beneficial.

Furthermore, while \RCE instances are designed for interconnecting with each other, and for actively managing other tools, it provides relatively few options for integrating itself into existing software ecosystems.
To address this, we intend to provide explicit APIs that can then be invoked by other software, e.g. for starting and querying workflow executions, accessing the data management, or managing network connections.

\subsection{User Management and Access Control}
\label{sec:future-work:user-management}

Currently, \RCE only provides a basic mechanism for controlling access to published tools, the "authorization groups" described in Section~\ref{sec:practice:authorization-tool-access}.
The common security boundary at the moment is the access to \RCE networks themselves; internally, there are few restrictions so far.
In particular, \RCE does not feature detailed mechanisms for restricting the access to workflow data, more fine-granular permissions for tool execution, or the monitoring of workflows yet.
Moreover, the access groups for controlling tool access are managed by individual \RCE users, i.e., the management of user groups is decentralized.

From an IT administrator's perspective, a centralized approach would fit better into the IT structures of most organizations.
This particularly involves central control and revocation of access keys.
It is, e.g., critical for IT security that members leaving an organization are automatically prevented from accessing future work within the organization's network.
From a user's perspective, the approach currently implemented in \RCE has the benefit of allowing flexible and low-overhead creation of new groups.
A centralized approach, in contrast, would require organizational structures, responsibilities, and procedures for managing access control.

Permissible and practical methods for user authentication are largely determined by the policies of the companies using \RCE.
The variety of existing authentication schemes and approaches within different organizations complicates designing a common and generally applicable approach.
Thus, it is infeasible to develop a solution for user management and user authentication that serves all use cases encountered in practice at once.
As the networks and workflows built with \RCE continue to grow larger, however, a more centralized and sophisticated framework for access control and authentication will be indispensable.

We are looking to design and implement a framework that allows for lightweight user management, that can be sustainably integrated with existing solutions, and that allows for sufficiently granular control over information in coordination with future project partners.

\subsection{Cloud Computing}
\label{sec:future-work:cloud}

The increasing adoption of cloud computing in science and engineering offers new options and challenges for the deployment of scientific tools.
While \RCE does not yet have explicit support for cloud computing, its focus on distributed workflow management would be a natural fit for it.
With individual tools deployed in a cloud environment using standardized interfaces, \RCE could provide the infrastructure for orchestrating the execution of such tools in workflows.

However, the practical aspects regarding tool installations listed in Section~\ref{sec:practice:tool-services} still apply.
Thus, significant research and design work is still needed for safely moving real-world tools and workflows into cloud environments.

Finally, in addition to the topics mentioned in this section, we are working closely with known users of \RCE to determine and anticipate current and future requirements towards \MBSE support in general, and particular needs regarding the software itself.
This work helps us prioritize and steer future development.

\section{Conclusion}
\label{sec:conclusion}

In this work we have presented \RCE, an application developed at \DLR that allows its users to construct and automatically execute tool chains comprised of both off-the-shelf and tailor-made software tools.
This aids engineers in constructing, analyzing, and optimizing the model constructed as part of \MBSE.
We have subsequently given an overview over \RCE's features and discussed the tradeoffs made in their practical implementation.
Moreover, we have highlighted projects in which \RCE has been used and given an outlook on future areas of development.

\RCE has enabled, supported, and simplified numerous research and development projects in a wide variety of fields over the years.
It is used in ongoing development projects, its adoption has significantly increased over the years, and there is great interest in a plethora of directions for future development on the side of users and project partners.
We are thus looking forward to continue development of \RCE to adapt it to new use cases and to evolve it in tandem with the state of the art in the wider field of application.

\bibliographystyle{splncs04}
\bibliography{aerospace2022-rce-for-mbse}

\end{document}

%% file: images/workflow-concept-evaluation.tex
\begin{tikzpicture}[thick,xscale=1.25,->,>=stealth,baseline]
  \node[draw, align=center] (simulation) at (0,0) {Scenario\\Simulation};
  \node[draw, align=center] (econ-evaluation) at (2,1) {Economic\\Evaluation};
  \node[draw, align=center] (perf-evaluation) at (2,0) {Performance\\Evaluation};
  \node[draw, align=center] (ecol-evaluation) at (2,-1) {Ecological\\Evaluation};
  \node[draw, align=center] (comb-evaluation) at (4,0) {Evaluation\\Summary};

  \path
    ($(simulation.west) - (.5,0)$) edge (simulation.west)
    (simulation)
      edge[out=60,in=180] (econ-evaluation)
      edge (perf-evaluation.west)
      edge[out=-60,in=180] (ecol-evaluation)
    (econ-evaluation.east) edge[out=0,in=120] (comb-evaluation)
    (perf-evaluation.east) edge (comb-evaluation)
    (ecol-evaluation.east) edge[out=0,in=-120] (comb-evaluation)
    (comb-evaluation.east) edge ($(comb-evaluation.east) + (.5,0)$);

\end{tikzpicture}

%% file: images/workflow-concept-main.tex
\begin{tikzpicture}[thick,xscale=1.25,->,>=stealth,baseline]
  \node[draw,align=center] (opt) at (0,0) {Model\\Generator};

  \node[draw,align=center] (scen-loader) at (2, 1) {Scenario \\ Loader};
  \node[draw,align=center,fill=gray!20] (scen-eval-1) at (4, 1) {Scenario \\ Evaluation};

  \node[draw,align=center] (scen-gen) at (2, -1) {Scenario \\ Generator};
  \node[draw,align=center,fill=gray!20] (scen-eval-2) at (4, -1) {Scenario \\ Evaluation};

  \path
    ($(opt) - (1,0)$) edge (opt);
  \path[draw,rounded corners] (opt) -| (scen-eval-1);
  \path[draw,rounded corners] (opt) -| (scen-eval-2);
  \path
    (scen-loader) edge (scen-eval-1);
  \path[transform canvas={yshift={.05cm}}]
    (scen-gen) edge (scen-eval-2);
  \path[transform canvas={yshift={-.05cm}}]
    (scen-eval-2) edge (scen-gen);

  \path[draw,rounded corners]
    (scen-eval-1.east) -| (5,1.75) -| (opt.north);
  \path[draw,rounded corners]
    (scen-eval-2.east) -| (5,1.75) -| (opt.north);

  \path[draw,rounded corners] (opt.south) |- (1, -1.75);
\end{tikzpicture}